\documentclass{aa}
\usepackage{graphicx}
\usepackage{float}
\usepackage{amsmath}
\usepackage{natbib}
\usepackage{subfigure}
\usepackage{hyperref}
\bibpunct{(}{)}{;}{a}{}{,}

\begin{document}
\title{Characteristics of magnetoacoustic sausage modes}
\author{A. R. Inglis
\and T. Van Doorsselaere
\and C. S. Brady
\and V. M. Nakariakov
 }

\institute{Centre for Fusion, Space and Astrophysics, Physics Department, University of Warwick, Coventry, CV4 7AL, UK}
\abstract{}
{We perform an advanced study of the fast magnetoacoustic sausage oscillations of coronal loops in the context of MHD coronal seismology to establish the dependence of the sausage mode period and cut-off wavenumber on the plasma-$\beta$ of the loop-filling plasma. A parametric study of the ratios for different harmonics of the mode is also carried out.}
{Full magnetohydrodynamic numerical simulations were performed using Lare2d, simulating hot, dense loops in a magnetic slab environment. The symmetric Epstein profile and a simple step-function profile were both used to model the density structure of the simulated loops. Analytical expressions for the cut-off wavenumber and the harmonic ratio between the second longitudinal harmonic and the fundamental were also examined.}
{It was established that the period of the global sausage mode is only very weakly dependent on the value of the plasma-$\beta$ inside a coronal loop, which justifies the application of this model to hot flaring loops. The cut-off wavenumber $k_c$ for the global mode was found to be dependent on both internal and external values of the plasma-$\beta$, again only weakly. By far the most important factor in this case was the value of the density contrast ratio between the loop and the surroundings. Finally, the deviation of the harmonic ratio $P_1/2P_2$ from the ideal non-dispersive case was shown to be considerable at low $k$, again strongly dependent on plasma density. Quantifying the behaviour of the cut-off wavenumber and the harmonic ratio has significant applications to the field of coronal seismology.}{}

\date{Received 17 March 2009 /
Accepted 4 June 2009 }

\maketitle

\section{Introduction}

In recent years, improvements in instrumentation, data analysis techniques, and theory have led to confirmed observations of magnetoacoustic waves in the solar corona, particularly the longitudinal, kink, and sausage modes \citep[e.g.][]{2005LRSP....2....3N}. These waves attract attention mainly because of the possible role they play in coronal heating and as natural probes of coronal plasma. The latter concept gave rise to MHD coronal seismology, the novel approach to coronal plasma diagnostics by means of MHD waves.

\begin{figure}
\begin{center}
\subfigure[The global sausage mode]{\label{fundamental}\includegraphics[scale=0.3]{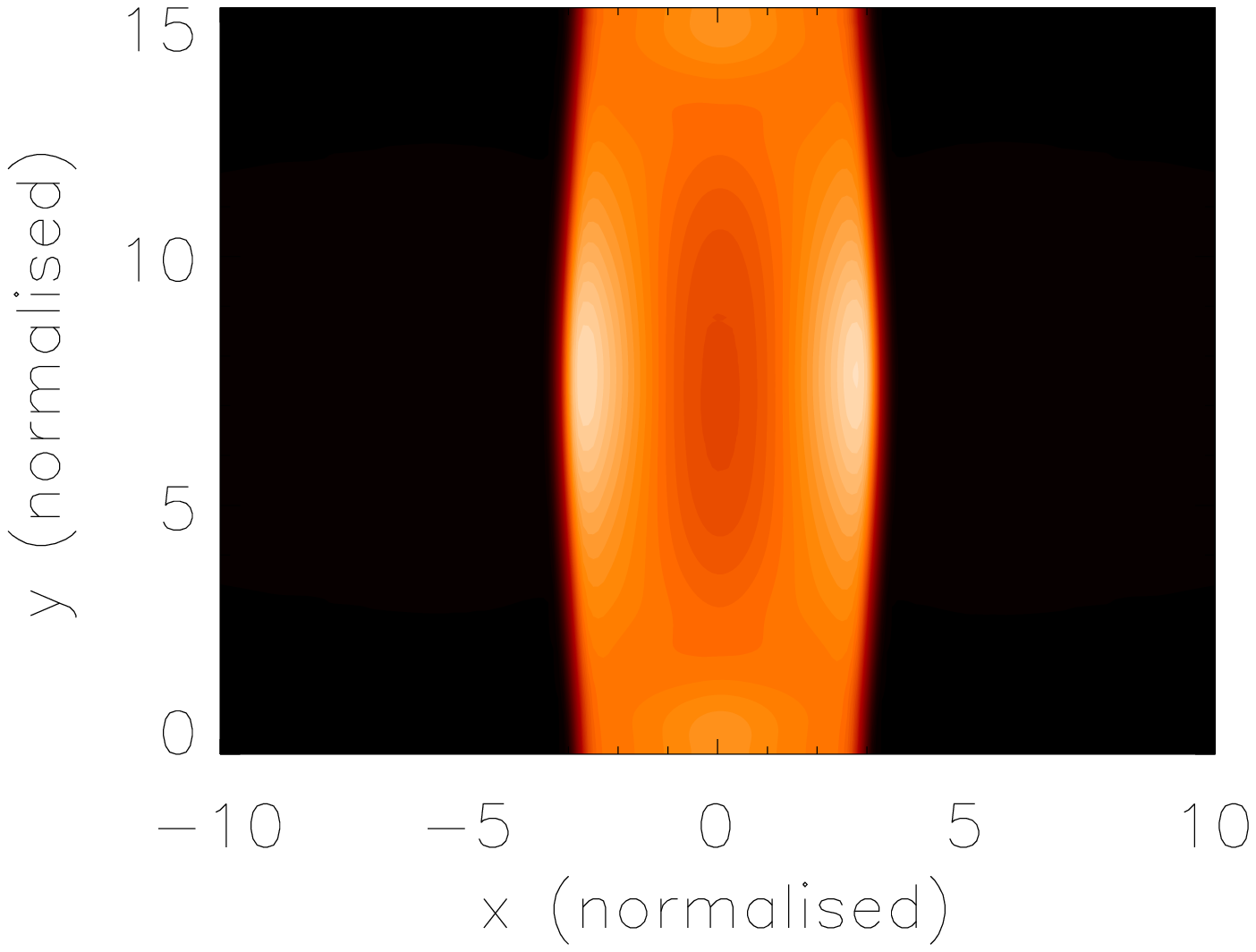}}
\subfigure[The second (longitudinal) harmonic]{\label{second_harmonic}\includegraphics[scale=0.3]{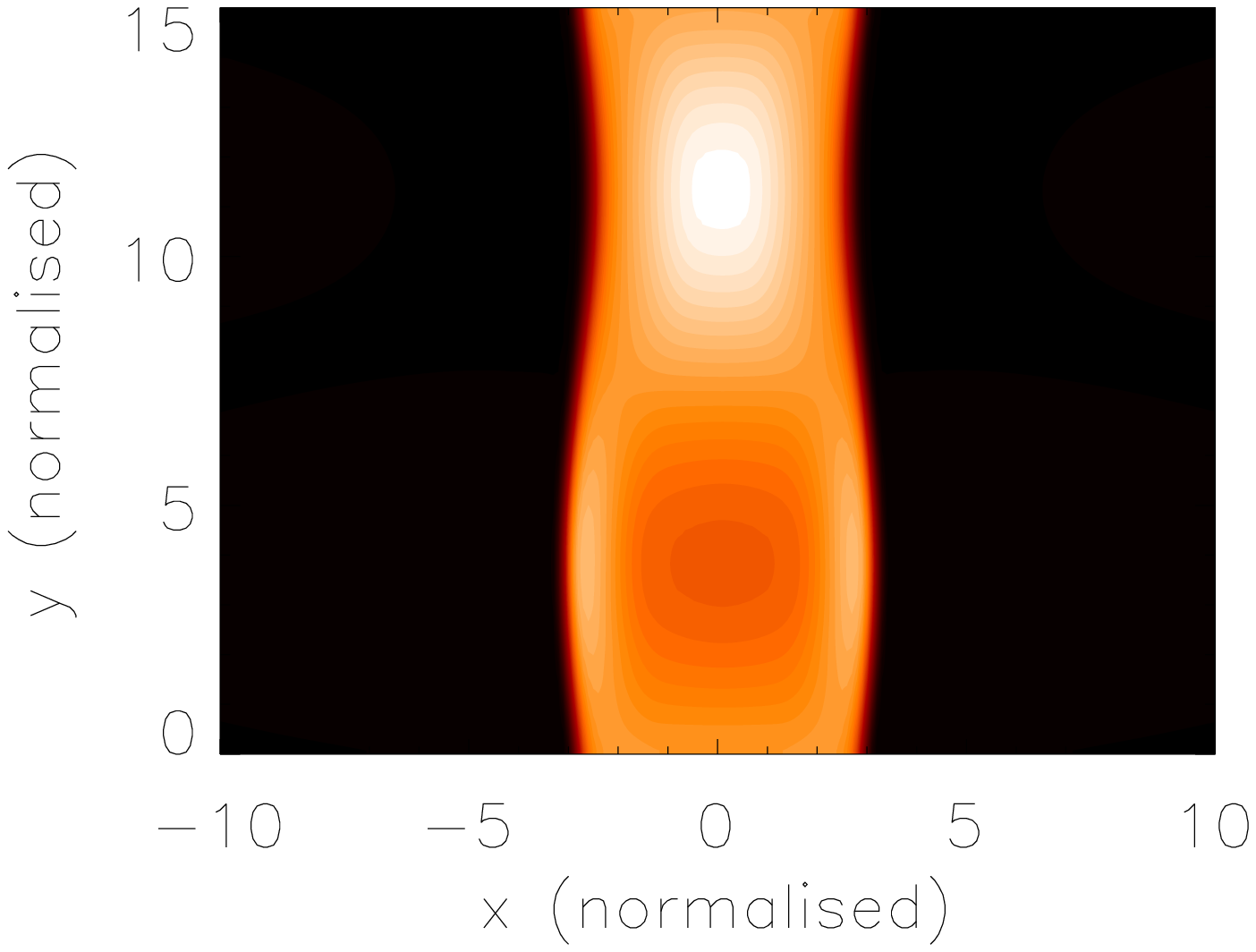}}
\end{center}
\caption{Illustrations of a) The global sausage mode and b) The second (longitudinal) harmonic of the sausage mode, shown as filled density contours. In both cases the perturbation is symmetric about the loop axis, which remains unperturbed.}
\label{sausage_mode}
\end{figure}

The sausage mode (also known as peristaltic or $m=0$ mode) is a symmetric perturbation of the loop minor radius, in the form of periodic broadening and narrowing of a plasma tube (see Figure \ref{sausage_mode}). This mode, with typical periods in the range 5-30~s, is believed to be detected in the microwave and hard X-ray emission associated with flaring solar loops \citep[e.g.][]{2003A&A...412L...7N,Melnikov,Inglis}, and also in $H_\alpha$ emission from cool, post-flare loops \citep{2008MNRAS.388.1899S}. The sausage mode is essentially compressible, with the density perturbations in phase with the perturbations of the magnetic field and in anti-phase with the perturbations of the loop minor radius. This mode is a robust collective disturbance of a plasma structure, which is practically insensitive to small scale irregularities of the plasma \citep{2007SoPh..246..165P}. The sausage mode can be manifest in both standing and travelling form. It is highly dispersive, with a pronounced dependence of the phase and group speeds upon the wave number. In the standing regime, this makes the spectrum of the resonant frequencies of different spatial harmonics non-equidistant. Long wavelength (in comparison with the minor radius of the oscillating loop) sausage modes are subject to a cut-off, in contrast with all other (kink and various fluting or ballooning) magnetoacoustic modes. Modes with wavelengths longer than the cut-off
are leaky, and their phase speed is slightly higher than the external Alfv\'en speed \citep{2007A&A...461.1149P}. For shorter wavelengths, sausage modes are trapped in the guiding structure, and have phase speeds in the range between the internal and external Alfv\'en speeds. This cut-off wavelength is dependent on the density contrast between the loop and its surroundings. As flaring loops are usually dense, thick and short, both trapped and leaky regimes can occur. The main properties of the sausage mode are derived from the straight cylinder or slab model. Clearly, the slab and the cylindrical cases are not identical, as has been discussed in \citet{2007AstL...33..706K}. In particular, \citet{2006ApJ...650L..91T} demonstrated the important difference in the wave leaking between these geometries. However, the difference in geometry should not be the principal difference in the resonant periods of sausage modes, as the behaviour of this mode is known to be almost
identical in the slab and cylinder geometries in the trapped regime. Future study of the sausage mode
of a plasma cylinder will provide a final answer, but is out of the scope of this paper.


Because of its compressibility, the sausage mode can be readily seen in microwave, EUV and soft X-ray bands, provided sufficient time resolution is available. Also, this mode is capable of modulating the population of nonthermal electrons via changes in the magnetic mirror ratio in the flaring loop, leading to periodic precipitation of electrons to the loop footpoint. This causes modulation of the emission in hard X-ray and white light because of bremsstrahlung \citep{1982SvAL....8..132Z}.

The seismological potential of the sausage mode is connected with estimating the magnetic field outside the oscillating plasma structure \citep{2007A&A...461.1149P}, and with determining the transverse profile of the plasma density in the flaring loop \citep{2004MNRAS.349..705N}. Also, in the case when the microwave spectrum is generated by the gyrosynchrotron mechanism and is spatially resolved, this mode provides a promising basis for combined MHD-microwave diagnostics of flaring plasmas. As the sausage mode often causes pronounced modulation of the emission light curve, it also opens up possibilities for the diagnostics of stellar coronae \citep{2008JPhCS.118a2038N}.

One popular topic of MHD coronal seismology is the use of the ratios of global (or fundamental) modes and their second harmonics (variously referred to as the $P_1/2P_2$-ratio or the $P_2/P_1$ ratio) for determining the longitudinal profile of the plasma parameters in the oscillating loop. \cite{2005ApJ...624L..57A} noticed that for the kink mode the discrepancy between the observed $P_1/2P_2$ ratio and that theoretically calculated with the use of the straight cylinder model could be attributed to the variation of the plasma density along the loop, e.g. because of gravitational stratification. There have been several attempts to account for this stratification in the theoretical model \citep[see][]{2006A&A...457.1059D,2006A&A...457..707D,2008A&A...481..819M,2008A&A...486.1015V}, and also to extend this study to longitudinal modes \citep[e.g.][]{2006A&A...458..975D}. One of the difficulties with the full scale implementation of the $P_1/2P_2$-based technique in MHD coronal seismology is that the higher harmonics of the kink mode are rarely detected because of insufficient temporal resolution of solar EUV imagers.

As the sausage mode is usually observed with the instruments which have very high time resolution (e.g. 0.1 or 1 s in the case of the Nobeyama Radioheliograph), it is possible to detect several harmonics of this mode \citep{2003A&A...412L...7N,Melnikov}.
However, as in contrast with the kink mode the sausage mode is highly dispersive, the identification of the sausage mode harmonic number is a non-trivial task. This is partly due to a significant departure of the $P_1/2P_2$-ratio from unity because of the dispersive modification of the phase speed even in the absence of longitudinal stratification. It becomes especially difficult when spatial resolution is not available. Hence there is need for detailed theoretical modelling of various regimes of the sausage mode dynamics.

It is not likely that in flaring loops the sausage mode is affected by gravitational stratification, as in such loops the density scale height is large because of the high temperature. The effect of the variable cross-section of the flaring loop on the resonant periods of the standing
sausage mode was recently considered by \citet{Pascoe2009}. It was established that the sausage mode continued to be supported despite variations in the loop cross-section. It was also shown that the characteristic periods of the global mode and higher harmonics were effected, but weakly so. A similarly small variation in parameters was found when investigating the effects of twist in loops \citep{2006SoPh..238...41E}.
However, the high temperatures and densities of flaring loops also lead to the increase in the plasma-$\beta$ inside the loops \citep{2001ApJ...557..326S}. This can cause a significant deviation of the resonant periods and cut-offs from the results obtained in the zero-$\beta$ approximation studied in \citet{2003A&A...409..325C,2007A&A...461.1149P}.

The purpose of this study is to establish the effect of finite plasma-$\beta$ on the resonant periods of the sausage mode by full numerical MHD modelling, and also to investigate the behaviour of the harmonic ratio for the sausage mode. The paper is organised as follows: in Section~2 we describe the standard theoretical approach to modelling wave modes in a magnetic slab geometry. Section 3 details the numerical methods used for the modelling in this paper. In Section 4 we present our results, which are then discussed in more detail in Section 5.

\section{Sausage modes in a magnetic slab}

Following the standard approach, we model magnetoacoustic oscillations in a coronal loop as perturbations of a field aligned enhancement in plasma density. The magnetic field is taken to be in the y-direction. The density profile is uniform in the y-direction. In the transverse direction the density is given by the profile \citep{1995SoPh..159..399N}

\begin{equation}
 \rho(x) = \left( \rho_0 - \rho_e \right) \mathrm{sech}^{2} \left[ \left( \frac{x}{a} \right)^n \right] + \rho_e
 \label{geps}
\end{equation}
where $\rho_e$ is the external density far from the loop, $\rho_0$ is the density at the loop apex and $a$ is a characteristic length governing the width of the loop. Here, $n=1$ corresponds to the symmetric Epstein profile (see Figure~\ref{density_b_temp}) which allows for analytical treatment. Increasing values of $n$ lead to a steeper profile, meaning that with a large enough exponent ($n\to\infty$) the profile given in Equation (\ref{geps}) can be considered as an approximate step function, also with a known analytical solution for fast magnetoacoustic modes \citep{1982SoPh...76..239E}. Most importantly, the use of this profile avoids steep gradients that may lead to strong artificial shocks in numerical simulations.

The equilibrium condition is the total pressure balance between the internal and external plasmas. This may be written as \citep{1982SoPh...76..239E}

\begin{equation}
 \frac{\rho_e}{\rho_0} = \frac{2 C_\mathrm{s0}^{2} + \gamma C_\mathrm{A0}^{2}}{2 C_\mathrm{se}^{2} + \gamma C_\mathrm{Ae}^{2}},
\label{pressure_balance}
\end{equation}
where $\gamma= \frac{5}{3}$ is the adiabatic index, $C_\mathrm{A0}$ and $C_\mathrm{Ae}$ are the internal and external Alfv\'en speeds, and $C_\mathrm{s0}$ and $C_\mathrm{se}$ are the internal and external sound speeds.

For a symmetric Epstein profile, the solutions describing the perturbation of the transverse velocity component in the sausage mode are given by

\begin{equation}
 U(x) = \frac{\mathrm{sinh}(x/a)} {\mathrm{cosh}^{\lambda}(x/a)},
\end{equation}
where

\begin{equation}
 \lambda = \frac{\left( C_\mathrm{Ae}^{2} - C_\mathrm{p}^{2} \right) ^{1/2} ka} {C_\mathrm{Ae}} + 1,
\end{equation}
where $k$ is the wavenumber and the phase speed $C_\mathrm{p}=\omega/k$ is in the range $C_\mathrm{A0}<C_\mathrm{p}<C_\mathrm{Ae}$ \citep{2003A&A...409..325C}.

Since $C_\mathrm{p}$  is typically a large fraction of $C_\mathrm{Ae}$ (for trapped modes) $\lambda$ is usually of order 1. The phase speed $C_\mathrm{p}$ can be worked out analytically from the dispersion relation

\begin{equation}
 \frac{ka}{C_\mathrm{A0}^{2}} \left( C_\mathrm{p}^{2} - C_\mathrm{A0}^{2} \right) - \frac{2}{ka} = \frac{3}{C_\mathrm{Ae}} \sqrt{C_\mathrm{Ae}^{2} - C_\mathrm{p}^{2}} .
\label{d_relation}
\end{equation}
This equation allows for the exact analytical solution for the phase speed $C_\mathrm{p}$. Unfortunately, this solution is only valid in the case of zero-$\beta$ plasmas.

For a realistic flaring loop the situation is likely to be more complex. A perfect pressure balance, for example, may not hold. Nor are the background plasma parameters likely to be uniform, or the amplitudes of oscillation sufficiently small to be considered linear. However, the global sausage mode is a very robust perturbation, and it remains desirable to examine its properties in a more idealised case before considering higher order effects. It is also likely that the choice of geometry is relatively unimportant for the MHD sausage mode. Part of the reason for this is its simplicity; a symmetric density perturbation about the loop axis independent of the azimuthal angle. This is in contrast to other MHD modes. In the following analysis, we simulate sausage modes using both a step-function density profile and a classical Epstein profile, noting the rationale in each case.

\section{Numerical model}

The numerical simulations of MHD sausage modes presented here were performed in slab geometry using Lare2d \citep{2001JCoPh.171..151A}. The code solves the MHD equations in normalised form. The defined normalisation constants used in this paper are:

\begin{displaymath}
L_0 = 1~ \mathrm{Mm},
\end{displaymath}
\begin{displaymath}
B_0 = 1 \times 10^{-3}~ \mathrm{T} (10~ \mathrm{G}),
\end{displaymath}
\begin{displaymath}
\rho_0 = 7.95 \times 10^{-13}~ \mathrm{kgm^{-3}}.
\end{displaymath}

This leads to the additional quantities:
\begin{displaymath}
t_0 =  1~ \mathrm{s},
\end{displaymath}
\begin{displaymath}
v_0 = 1\times 10^{6}~ \mathrm{ms^{-1}},
\end{displaymath}
\begin{displaymath}
T_0 = 6\times 10^{7}~\mathrm{K}.
\end{displaymath}

The normalised MHD equations may then be written as

\begin{equation}
 \frac{D\rho}{Dt} = -\rho \nabla \cdot \vec{\upsilon},
\end{equation}

\begin{equation}
 \rho \frac{D \vec{\upsilon}}{Dt} = \left( \nabla \times \vec{B} \right) \times \vec{B} - \nabla P,
\end{equation}

\begin{equation}
 \frac{D\vec{B}}{Dt} = \left( \vec{B} \cdot \nabla \right) \vec{\upsilon} - \vec{B} \left( \nabla \cdot \vec{\upsilon} \right),
\end{equation}

\begin{equation}
 \rho \frac{D \epsilon}{Dt} = -P\nabla \cdot \vec{\upsilon},
\end{equation}
where $\vec{\upsilon}$ is the plasma velocity, $P$ is the gas pressure, $\vec{B}$ is the magnetic field, $\rho$ is the mass density and $\epsilon = P/\rho(\gamma - 1)$ is the internal energy density.

\begin{figure}
 \begin{center}
   \includegraphics[width=8cm]{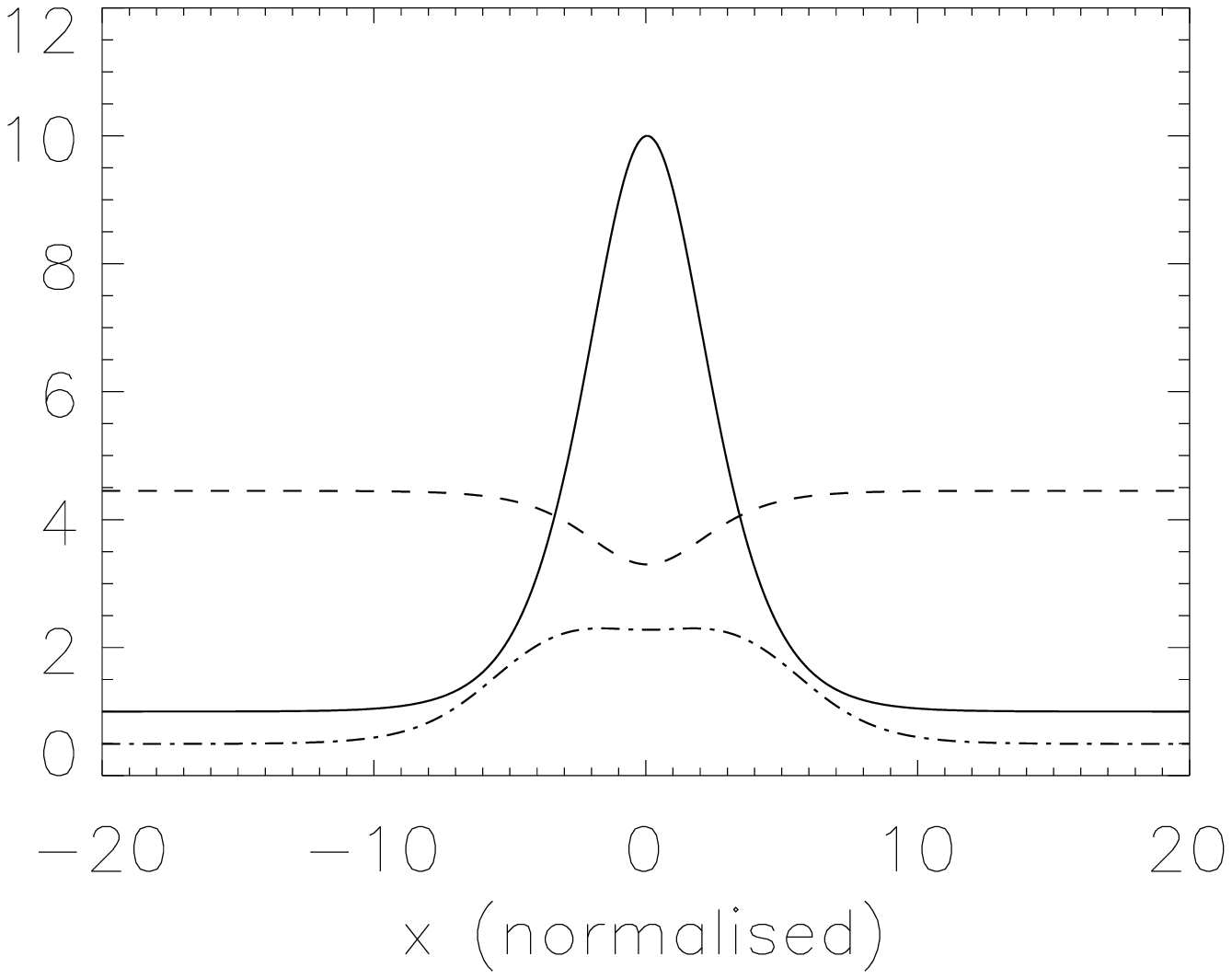}
 \includegraphics[width=8cm]{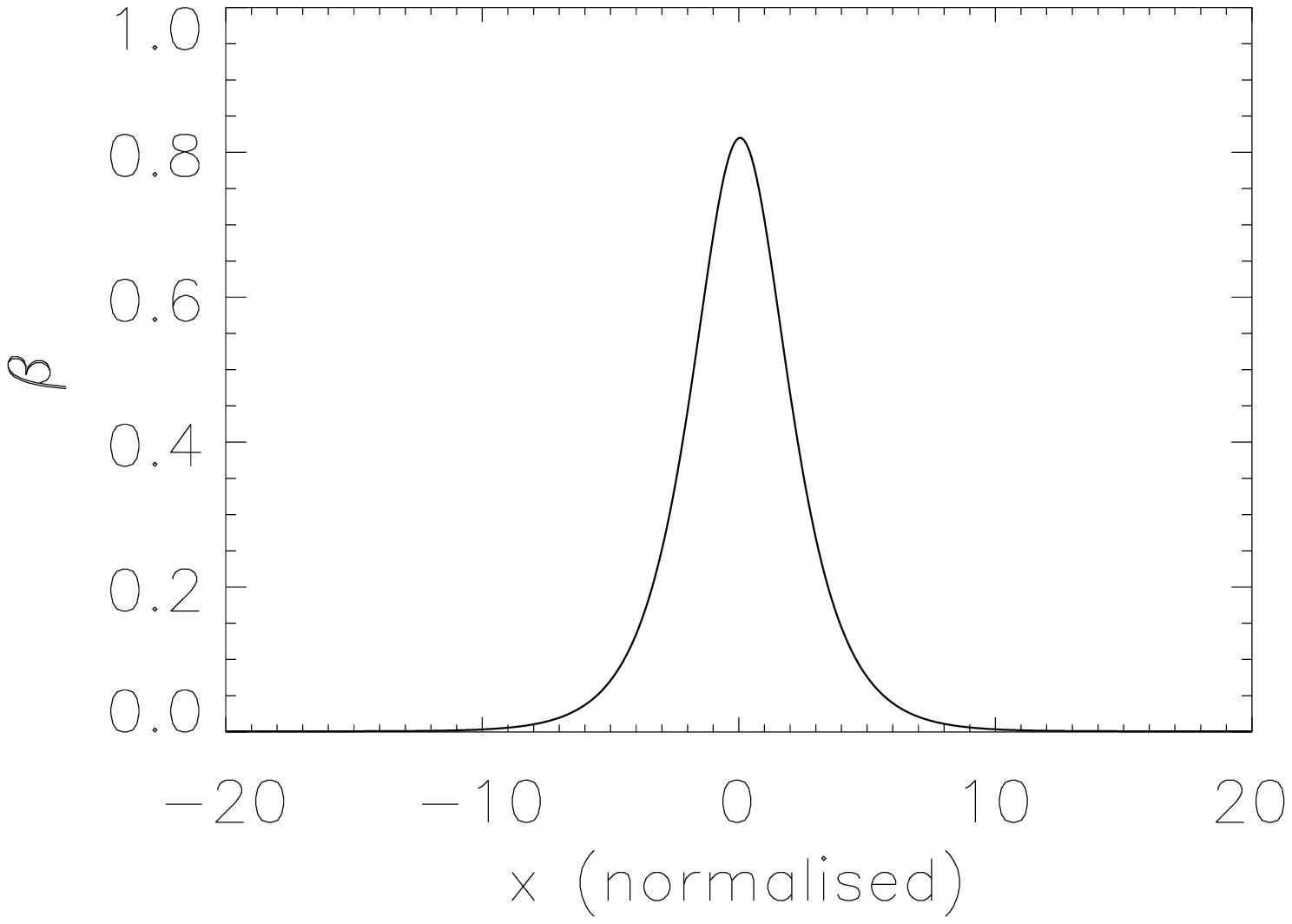}
\caption{Top panel: plot of the normalised initial parameters as a function of the transverse coordinate. The solid line is the density, the dashed line is the magnetic field $B_y$, and the dot-dashed line is the temperature. The temperature has been multiplied by a factor of 5 for clarity. Bottom panel: the $\beta$ profile resulting from the parameters shown in the top panel.}
\label{density_b_temp}
 \end{center}

\end{figure}

An initial structure in equilibrium was set up as described in Section 2. In order to control the value of $\beta$, the magnetic field was reduced inside the loop as shown in Figure \ref{density_b_temp}. All of the values in Figure \ref{density_b_temp} are normalised. The internal density of the loop, $\rho_0$, is also fixed. The internal gas pressure is left as a free parameter in order to maintain total pressure balance. As a result the plasma temperature varies with the internal gas pressure. Consequently, the simulated plasma structures are both hotter and denser than the surroundings, consistent with the known properties of physical coronal loops.

A convenient way to modify the equilibrium is to adjust the profile of the magnetic field, keeping all other parameters the same. This is the approach that will be used throughout this paper. This would allow us to consider the important case of flaring loops filled in with a hot and dense, high-$\beta$ plasma. 

In the following, the values of plasma-$\beta$ in the centre of the loop and at infinity are referred to as $\beta_0$ and $\beta_e$, respectively. To simulate a sausage mode we set up an initial transverse velocity perturbation of the form

\begin{equation}
 v_x = A U(x) \sin (ky),
\label{velocity}
\end{equation}
where $A$ is an amplitude and $x$ and $y$ are the coordinates across and along the slab respectively. This velocity distribution disturbs the loop at its centre and ensures that a global standing mode is developed. To avoid nonlinear effects the amplitude of the perturbation must be small compared to the equilibrium parameters. We choose $A$ = 0.002 in all cases, resulting in perturbations which are very small fractions of the equilibrium magnetic field and density (typically $\approx$ 0.1\% of the internal density for example). As described in Section 2, the magnetic field is aligned in the y-direction, as is the loop itself. Reflective boundaries were applied in the y-direction, while open boundaries were applied in the transverse direction. 

Theoretically, any excitation of a mode in this way would lead to the excitation of higher harmonics \citep{2005A&A...441..371T}. However, the excitation we choose is always sufficiently close to the eigenmode that the amplitude of these harmonics is very small with respect to the fundamental oscillation, and they are quickly damped. Thus, by running the simulation for sufficient time, the dominance of the desired mode is ensured \citep{2007A&A...461.1149P}.

To measure the periods of oscillation, the perturbation of the plasma parameters at the relevant anti-node are studied. For the global sausage mode this is at the loop centre. The period is determined by applying a Fourier transform to the time series data from this point, thus the uncertainty in the measurement is connected with the half-width of the peak in the Fourier spectrum.

\section{Results}
\subsection{The period of the sausage mode}

According to \citet{2003A&A...412L...7N}, the period of the global sausage mode near the wavenumber cut-off in the trapped regime is given by

\begin{equation}
 P \approx \frac{2L}{C_\mathrm{p}},
\end{equation}
where $L$ is the loop length and $C_\mathrm{p}$ is the phase speed. More recently, this relation was shown to be a valid approximation in the leaky regime \citep{2007A&A...461.1149P} and in a loop with variable cross-section \citet{Pascoe2009}. However, in both cases this relation was tested for very low values of $\beta_0$. It is unclear whether or not a significant value of $\beta_0$ would have a strong effect on the period of the mode. As a first step therefore, we reproduce the result of \citet{2007A&A...461.1149P} showing the dependence of the period on the loop length $L$, but this time for a variety of $\beta_0$ values. This is achieved by simulating sausage oscillations in a loop with an Epstein density profile and a density contrast ratio of 10.

\begin{figure}
 \begin{center}
   \includegraphics[width=8cm]{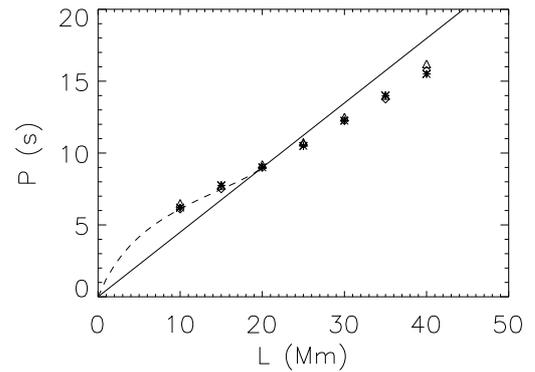}
  \caption{Period of the global sausage mode as a function of loop length $L$, for a number of $\beta_0$ values. The stars correspond to $\beta_0$=0.005, the diamonds to $\beta_0$=0.08 and the triangles to $\beta_0$=0.82. The solid line corresponds to $2L/C_\mathrm{Ae}$, while the dashed line is the analytical solution to the dispersion relation for an Epstein density profile. The density contrast ratio is 10.}
 \label{pascoe_fig6}
 \end{center}

\end{figure}

\begin{figure}
 \begin{center}
   \includegraphics[width=8cm]{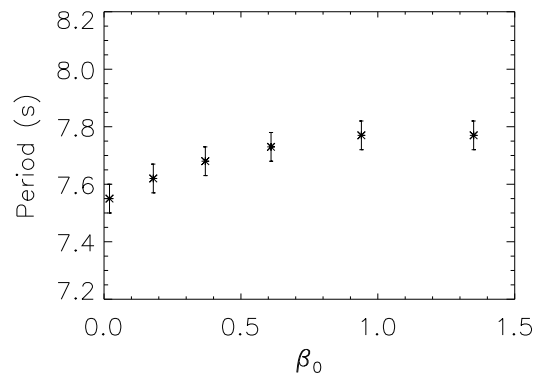}
\caption{The sausage mode period dependence on $\beta_0$ for a trapped mode in a loop of length $L$=15~Mm. The density contrast ratio is 10.}
\label{period_vs_beta}
 \end{center}

\end{figure}

The results shown in Figure \ref{pascoe_fig6} are in excellent agreement with those presented in \citet{2007A&A...461.1149P}. Above approximately 20 Mm the mode becomes leaky and the phase speed is slightly higher than the external Alfven speed. Furthermore, the near-independence of the oscillation period on $\beta_0$ is clear. As a further illustration, we investigate the period dependence on $\beta_0$ at a fixed length $L$. The length of the loop was set at $L$=15 Mm - corresponding to trapped modes - and the value of $\beta_0$ was varied as described in Section 3. $\beta_e$ was kept constant and small throughout. The result is displayed in Figure \ref{period_vs_beta}.

\begin{figure}
 \begin{center}
  \includegraphics[width=8cm]{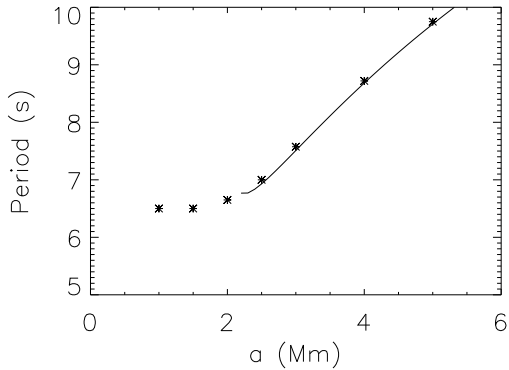}
\caption{The sausage mode period dependence on width $a$ for a loop of fixed length $L$=15~Mm, where the density contrast ratio is 10. The solid line corresponds to the solution of the dispersion relation for an Epstein profile.}
\label{period_vs_width}
 \end{center}

\end{figure}

The period of the sausage mode should also depend on the loop width $a$. This is due to dispersion, since a change in $a$ essentially corresponds to movement along the dispersion curve derived from Equation \ref{d_relation} and hence a change in phase speed $C_p$. To illustrate this, we performed numerical simulations of a sausage mode in a loop of fixed length $L$=15~Mm while varying the width of the Epstein profile. The results are illustrated in Figure \ref{period_vs_width}. As expected, the period increases as a function of $a$ within the trapped regime, whereas for leaky modes - corresponding to $a < 2.2$ Mm - the period tends to a constant value. This is a consequence of the phase speed remaining close to the external Alfven speed in this regime. 

It is clear from these results that, although there is a slight increase in the period with respect to $\beta_0$, the variation is very small (less than 5\%) compared to the effect of other parameters such as loop length. Thus we may say that the period of the sausage mode is very stable with respect to the value of $\beta_0$.

\subsection{The second longitudinal harmonic}

An often-neglected aspect of sausage mode behaviour is the relationship between the periods of separate harmonics. For dispersionless standing modes in a resonator, the frequency of the $N$-th harmonic is simply $N$ times that of the fundamental. Being strongly dispersive, the sausage mode harmonics deviate significantly from this pattern.

\begin{figure}
 \begin{center}
\includegraphics[width=8cm]{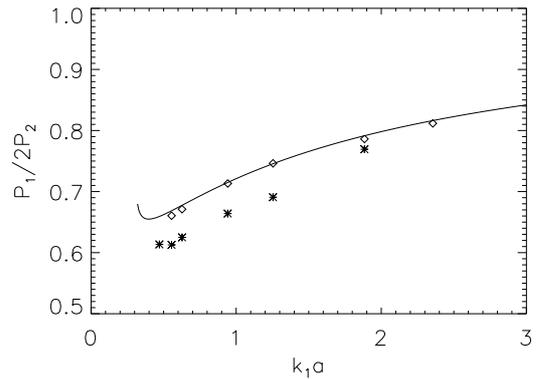}
\includegraphics[width=8cm]{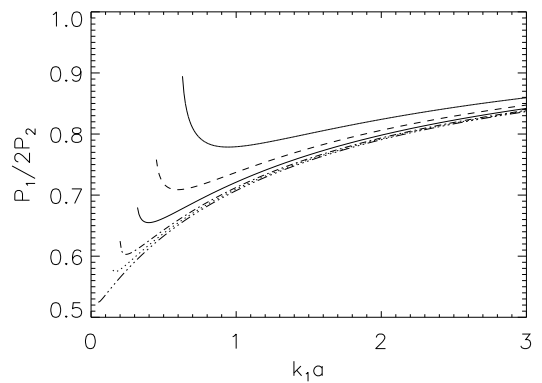}
\caption{Top panel: The ratio $P_1/2P_2$ observed via full numerical simulation for the sausage mode as a function of the longitudinal wave number $k_1$ for both the step-function (stars) and Epstein profile (diamonds) geometries. The density contrast ratio is 20 and $\beta_0$=0.08. The solid line is the theoretical curve derived from the dispersion relation for the Epstein profile, also using a density contrast ratio of 20.
Bottom panel: Illustration of the dependence of the analytical solution on the density contrast ratio. The solution is shown for contrast ratios of 5 (upper solid line), 10 (dashed line), 20 (solid line), 50 (dot-dashed line), 100 (dotted line) and 1000 (triple-dot-dashed line).}
\label{period_ratio}
\end{center}
\end{figure}

The true behaviour of the harmonic ratio is explored for the fundamental and second harmonic modes. For a given value of $k_1$ (the wavenumber corresponding to the fundamental mode), the period ratio $P_1/2P_2$ is measured from numerical simulations. By varying the loop length $L$, we control the wavenumber $k_1$ and subsequently measure the period ratio as a function of $k_1$. Figure \ref{period_ratio} displays the results.

For non-dispersive waves, the expected ratio of $P_1/2P_2$ is 1. It is clear that the sausage mode harmonics do not follow this pattern. For large values of $k_1$ corresponding to the uniform plasma (and hence dispersionless) limit, the ratio does indeed tend towards the ideal value. However, for smaller $k_1$, the period of the second harmonic becomes a much greater fraction of the fundamental period, as illustrated in both panels of Figure \ref{period_ratio}. The top panel shows the quantitative difference between simulations of an Epstein profile (diamonds) and a step-function profile (stars). Although this deviation is noticeable at low $k_1$ values, the overall behaviour remains the same, as expected. The solid line shows the theoretical curve for an Epstein profile derived from the dispersion relation given in Equation (\ref{d_relation}), corresponding to the zero-$\beta$ case. The agreement between the theoretical and numerical results is impressive.
The bottom panel of Figure \ref{period_ratio} shows the same theoretical curve for a range of density contrast ratios. Clearly there is a strong effect on the behaviour of the period ratio; the denser the loop, the greater the deviation from the ideal, non-dispersive case. However, the range for the period ratio in all cases lies between 0.5 and 1. The turnover of the period ratio curve seen at low $k_1$ is a product of being very close to the cut-off wavenumber $k_c$.

In coronal seismology, the ratios of observed oscillatory periods - say, in a coronal loop or solar flare - are often used to make judgements about the underlying MHD modes. Figure \ref{period_ratio} provides a powerful illustration of the dispersive nature of sausage mode harmonics, and that for entirely realistic values of $k$ it is not expected that the harmonic ratio be close to 1 regardless of any non-uniformity in the longitudinal direction.

\subsection{The cut-off wavenumber}
\label{section_cutoff}
According to \citet{1995SoPh..159..213N}, the cutoff wavenumber for a sausage mode in a magnetic slab with a step function profile of finite-$\beta$ plasma is given by

\begin{equation}
k_{c}a=\frac{\pi}{2}\sqrt{-\frac{\left(1+\frac{C_\mathrm{s0}^2}{C_\mathrm{A0}^2}\right)\left(\frac{C_\mathrm{s0}^2}{C_\mathrm{A0}^2+C_\mathrm{s0}^2}-\frac{C_\mathrm{Ae}^2}{C_\mathrm{A0}^2}\right)}{\left(1-\frac{C_\mathrm{Ae}^2}{C_\mathrm{A0}^2}\right)\left(\frac{C_\mathrm{s0}^2}{C_\mathrm{A0}^2}-\frac{C_\mathrm{Ae}^2}{C_\mathrm{A0}^2}\right)}}.\label{nakariakov1995}
\end{equation}








An analytical expression for the wavenumber cut-off is not readily obtainable in the finite-$\beta$ case for the symmetric Epstein profile.
The available expression (Pascoe et al. 2007) is only valid when $\beta$=0. Therefore for consistency all analysis of the
cut-off wavenumber is carried out in step-function geometry.

In order to investigate the dependence of this cut-off wavelength on the plasma-$\beta$, we write the sound speed as $C_\mathrm{s}^2/C_\mathrm{A}^2=\gamma \beta/2$. Equation \ref{nakariakov1995} then becomes

\begin{eqnarray}k_{c}a&=&\frac{\pi}{2}\sqrt{-\frac{\left(1-\frac{C_\mathrm{Ae}^2}{C_\mathrm{A0}^2}\right)\frac{\gamma \beta_\mathrm{0}}{2}-\frac{C_\mathrm{Ae}^2}{C_\mathrm{A0}^2}}{\left(1-\frac{C_\mathrm{Ae}^2}{C_\mathrm{A0}^2}\right)\left(\frac{\gamma \beta_\mathrm{0}}{2}-\frac{C_\mathrm{Ae}^2}{C_\mathrm{A0}^2}\right)}}\\
 &=&\frac{\pi}{2}\sqrt{-1+\frac{\left(\frac{C_\mathrm{Ae}^2}{C_\mathrm{A0}^2}\right)^2}{\left(1-\frac{C_\mathrm{Ae}^2}{C_\mathrm{A0}^2}\right)\left(\frac{\gamma \beta_\mathrm{0}}{2}-\frac{C_\mathrm{Ae}^2}{C_\mathrm{A0}^2}\right)}}.\label{eq:cutoffva}
\end{eqnarray}

Using the equation for magnetostatic pressure balance (Equation \ref{pressure_balance}), we can eliminate $C_\mathrm{Ae}/C_\mathrm{A0}$ and write the expression for the cut-off wavelength in terms of the density contrast $\zeta=\rho_\mathrm{0}/\rho_\mathrm{e}$ and the plasma-$\beta$. We find
\[\frac{C_\mathrm{Ae}^2}{C_\mathrm{A0}^2}=\zeta \left(1+\beta_\mathrm{0}\right)-\beta_\mathrm{e}.\]
Inserting this expression into Equation \ref{eq:cutoffva}, we obtain

\begin{equation}
 k_{c}a=\frac{\pi}{2}\sqrt{-1+\frac{\left(\zeta\beta_\mathrm{0}-\beta_\mathrm{e}+\zeta\right)^2}{\left(\zeta\beta_\mathrm{0}-\beta_\mathrm{e}+\zeta-1\right)\left((\zeta-\frac{\gamma}{2})\beta_\mathrm{0}-\beta_\mathrm{e}+\zeta\right)}
}.
\label{tom}
\end{equation}

\begin{figure}
 \begin{center}
  \includegraphics[width=8cm]{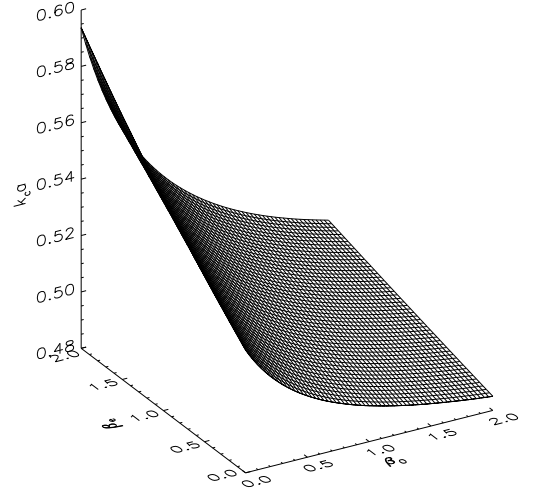}
\caption{Surface showing the dependence of the cut-off wavenumber on $\beta_0$ and $\beta_e$. The density contrast ratio is 10.}
\label{cutoff_surface2}
 \end{center}

\end{figure}

This dependence is shown in Figure \ref{cutoff_surface2}. For any value of $\beta_e$, $k_c a$ is seen to decrease as a function of increasing $\beta_0$. Realistically, in a coronal plasma the external $\beta_e$ is likely to be small. Thus we will concentrate on this region of the surface.

\begin{figure}
 \begin{center}
\includegraphics[width=8cm]{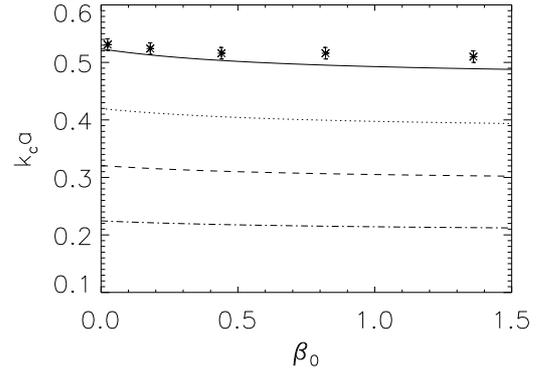}
\caption{Plot showing the dependence of $k_c a$ on $\beta_0$ for a fixed value of $\beta_e$, where $\beta_e$ = 0.01. The density contrast ratios used are 10 (solid line), 15 (dotted line), 25 (dashed line) and 50 (dot-dashed line). The overlayed datapoints are the results from full numerical simulations.}
\label{cutoff_fixed_be}
 \end{center}

\end{figure}

By comparison, we numerically evaluate Equation \ref{nakariakov1995} for increasing values of $\beta_0$, while maintaining a fixed and low $\beta_e$. We maintain fixed external values of $\rho$ and $B$. Inside the loop, the value of $B$ is decreased, increasing the value of $\beta_0$. For the structure to remain stable, the internal gas pressure must be left a free parameter to maintain total pressure balance. As before, the temperature is allowed to vary with the gas pressure, ensuring that the simulated loops are hot and dense. The results are shown in Figure \ref{cutoff_fixed_be}, where the solid line corresponds to a density contrast ratio of 10.




In both cases, the value of the cut-off wavenumber clearly decreases while $\beta_0$ is increased from zero to one. Figure \ref{cutoff_fixed_be} shows the same result for a number of different density contrast ratios, illustrating that the density has a strong impact on the confinement of the sausage mode. Again though for any starting - and fixed - value of the density contrast, the cut-off wavenumber will decrease as a function of $\beta_0$.



To further corroborate these results, we perform full numerical simulations of the global sausage mode for varying values of $\beta_0$. This is achieved by controlling the internal magnetic field as explained in Section 2. 

To measure the cut-off wavenumber itself, we simulate the global mode for varying $L$ (and hence $k$) and examine the time series of the loop anti-node as described in Section 3. Here the criterion for a mode to be leaky is that the amplitude of the oscillation inside the loop is damped. When far into the leaky regime, the damping effect is strong and evident \citep[see for example][]{2007A&A...461.1149P}. However, closer to the trapped regime it can be difficult to ascertain whether significant damping is taking place. This effect is mitigated by running the simulations for a sufficiently long time. However an exact determination of the cut-off wavenumber $k_c$ remains difficult.

The results of these MHD simulations are overlayed in Figure \ref{cutoff_fixed_be}. The simulations show the same general trend as the analytical and numerical solutions of Equation \ref{nakariakov1995}, although the agreement is not exact. The curve of the simulated result is noticeably shallower than the theoretical predictions. In part this is likely due to the difficulty in determining exactly the transition between the trapped and leaky regimes, a problem that is reflected in the size of the error bars shown for the numerical results.

\section{Conclusions}

Using numerical simulation and analytical study, we have carried out detailed analysis encompassing various aspects of MHD sausage mode behaviour. The key findings may be summarised as follows:

\begin{enumerate}
 \item In the finite-$\beta$ regime, the period of the global sausage mode was found to be determined by the length of the loop $L$, which confirmed previous findings of \citet{2003A&A...412L...7N}, \citet{2004ApJ...600..458A} and \citet{2007A&A...461.1149P}.
 \item In the trapped regime the period is also dependent on the loop width $a$ due to dispersion, but becomes weakly dependent in the leaky regime.
 \item The period of the global sausage mode is not significantly affected by finite values of $\beta_0$, with the variation remaining less than 5\% for $0 < \beta < 1$.
 \item The ratio between the second longitudinal harmonic and the fundamental, $P_1/2P_2$, varies between approximately 0.5 and 1 for `reasonable' values of $k_1$ and $\rho$, and is strongly dependent on both $k_1$ and $\rho$.
 \item The cut-off wavenumber for the global mode is a function of $\beta_0$, $\beta_e$ and the density contrast ratio $\zeta$. For a given density and $\beta_e$, the $k_c a$ is a decreasing function with respect to $\beta_0$.
\end{enumerate}

Analysis of the cut-off wavenumber $k_c$ reveals the limiting factors dictating whether an MHD mode remains trapped or becomes leaky. The unambiguous result is that the density contrast between the loop itself and the surroundings is the most important factor. For extremely dense loops, a sausage mode may remain trapped even for very long loop lengths. Conversely, for loops of modest density, it is clear that trapped modes may only be supported for comparatively short lengths.
The values of $\beta_e$ and $\beta_0$, although also factors, are secondary ones. It was shown in Section \ref{section_cutoff} that $k_c$ decreases as a function of $\beta_0$ when $\beta_e$ and $\zeta$ are treated as constants. In more physical terms, this means that the maximum length supporting trapped modes, $L_c$, is increased. However, this variation is clearly small when compared to the effect of the density contrast ratio $\zeta$ (see Figure \ref{cutoff_fixed_be}).

These results also have immediate practical applications in the context of MHD coronal seismology. For multi-periodic oscillations, an understanding of the relationship between sausage mode harmonics enables informed judgements about the nature of the observed mode.

For example, in \citet{2008MNRAS.388.1899S} a period ratio of $P_1/2P_2 \approx 0.83$ was reported, with the favoured interpretation being the fundamental and second harmonic of the sausage mode. From their estimates of the loop length ($L \approx$ 100 Mm) and our independent estimate of the loop width ($a \approx$ 8 Mm), it is possible to examine this conclusion. Given the values of $L$ and $a$, we estimate that $k_1 a \approx$ 0.25. It is clear from Figure \ref{period_ratio} that, for the period ratio reported this is very much in the leaky regime for the sausage mode. However, the data presented in \citet{2008MNRAS.388.1899S} does not show significant evidence of damping. Therefore, although it must be pointed out that the uncertainty in $k_1 a$ is considerable, it seems that the observed $P_1$/$2P_2$ ratio is an indication that another mode may be responsible.

We apply the same analysis to the event analysed in \citet{2003A&A...412L...7N}, where a fundamental and possible second harmonic was reported. In this case the observed period ratio is $P_1/2P_2 = 0.82 \pm 0.15$. Given the published estimates of the loop length ($L \approx$ 25 Mm) and width ($a \approx$ 6 Mm) we find that in this case $k_1 a \approx $ 0.75. Again, we can compare this with the results shown in Figure \ref{period_ratio}.  In this case, the observed period ratio and wavenumber are broadly consistent with a trapped fundamental and second harmonic of the sausage mode, provided the density contrast ratio is not too large, say 5 or perhaps even 10. Although this is not particularly large for a hot flaring loop, it is certainly possible and consistent with a trapped mode regime. Another possibility to consider is that the observed oscillations are actually triggered by a nearby non-flaring loop, which might have a lower density contrast than the flaring loop itself \citep{2006A&A...452..343N}. These two examples show how the results obtained here can be readily applied to observational examples of multi-period QPP.

Nevertheless, significant scope for improvement remains, both in observational and theoretical terms. In general, the uncertainties in both $P_1/2P_2$ and $k_1 a$ are substantial in observational studies of QPP. Additionally, as has been shown in Figure \ref{period_ratio}, the geometry of the theoretical model has a bearing on the quantitative values of the period ratio. Potentially important effects such as loop curvature and gravity have also been neglected in this study. Understanding the impact of these effects may prove an important next step in quantifying the characteristics of sausage modes.

\begin{acknowledgements}
 ARI would like to acknowledge the support of a STFC PhD studentship. TVD was supported by a Marie Curie Intra-European Fellowship within the 7th European Community Framework Programme. 
\end{acknowledgements}

\bibliographystyle{aa.bst}
\bibliography{12088.bib}

\end{document}